\documentclass[conference]{IEEEtran}

\usepackage{float}
\usepackage{amssymb}%
\usepackage{pifont}%
\newcommand{\cmark}{\ding{51}}%
\newcommand{\xmark}{\ding{55}}%
\usepackage{cite}

\usepackage{graphicx}
\usepackage{comment}

\ifCLASSINFOpdf
\else
\fi

\usepackage{algpseudocode}
\usepackage{algorithm}
\usepackage{algorithmicx}

\usepackage[hyphens]{url}
\usepackage{hyperref}
\usepackage{tabularx}
\newcolumntype{Y}{>{\centering\arraybackslash}X}

\begin{document}

\title{ExploitWP2Docker: a Platform for Automating the Generation of Vulnerable WordPress Environments for Cyber Ranges}

\author{\IEEEauthorblockN{1\textsuperscript{st} Francesco Caturano}
\IEEEauthorblockA{\textit{DIETI} \\
\textit{University of Naples “Federico II"}\\
Naples, Italy \\
francesco.caturano@unina.it}
\and
\IEEEauthorblockN{2\textsuperscript{nd} Nicola d'Ambrosio}
\IEEEauthorblockA{\textit{DIETI} \\
\textit{University of Naples “Federico II"}\\
Naples, Italy \\
nicola.dambrosio2@unina.it}
\and
\IEEEauthorblockN{3\textsuperscript{rd} Gaetano Perrone}
\IEEEauthorblockA{\textit{DIETI} \\
\textit{University of Naples “Federico II"}\\
Naples, Italy \\
gaetano.perrone@unina.it}
\and
\IEEEauthorblockN{4\textsuperscript{th} Luigi Previdente}
\IEEEauthorblockA{\textit{DIETI} \\
\textit{University of Naples “Federico II"}\\
Naples, Italy \\
lprevidente@gmail.com}
\and
\IEEEauthorblockN{5\textsuperscript{th} Simon Pietro Romano}
\IEEEauthorblockA{\textit{DIETI} \\
\textit{University of Naples “Federico II"}\\
Naples, Italy \\
spromano@unina.it}
}

\maketitle

\begin{abstract}
A cyber range is a realistic simulation of an organization's network infrastructure, commonly used for cyber security training purposes. It provides a safe environment to assess competencies in both offensive and defensive techniques.
An important step during the realization of a cyber range is the generation of vulnerable machines. This step is challenging and requires a laborious manual configuration. Several works aim to reduce this overhead, but the current state-of-the-art focuses on generating network services without considering the effort required to build vulnerable environments for web applications. A cyber range should represent a real system, and nowadays, almost all the companies develop their company site by using WordPress, a common Content Management System (CMS), which is also one of the most critical attackers' entry points.
The presented work proposes an approach to automatically create and configure vulnerable WordPress applications by using the information presented in public exploits. Our platform automatically extracts information from the most well-known publicly available exploit database in order to generate and configure vulnerable environments. The container-based virtualization is used to generate lightweight and easily deployable infrastructures. A final evaluation highlights promising results regarding the possibility of automating the generation of vulnerable environments through our approach.
\end{abstract}


\IEEEpeerreviewmaketitle

\section{Introduction}
The world is more and more interconnected, so the amount of sensitive data exchanged across the network is becoming increasingly important. As a result, cyberattacks are growing in terms of both numbers and level of sophistication.

In this scenario, companies pay more attention to cybersecurity to prevent cyberattacks and protect their business assets. According to Deloitte business research,  organizations will invest between $6$\% and $14$\% of their annual IT budget in cybersecurity~\cite{bernard_nicholson_2020}. Part of this budget will be devoted to educating inexperienced and insufficiently trained employees, who can represent a severe security risk.
In order to be effective, such training sessions must integrate complex hands-on cybersecurity exercises in addition to face-to-face theoretical lectures. Trainers can accomplish that goal through cyber ranges.

Cyber ranges are interactive, virtual representations of networks, systems, tools, and applications that enable learners to both practice and assess the acquired skills. These virtual scenarios provide a safe and legal environment to gain hands-on cyber skills, to learn secure development, and, more in general, to test an organization's security posture~\cite{cyberRangeNIST}. 
People can use cyber ranges for other purposes, too. For example, developers and system administrators might test systems and applications within cyber ranges and identify vulnerabilities in a controlled manner before going to production.

Unfortunately, the development and maintenance of a cyber range can become cumbersome. 
One of the most expensive tasks in terms of human effort is the realization of the vulnerable environment used to train. In fact, it is often done manually and requires a lot of time, effort, and skills. Several works try to address the issue by using information obtained through the definition of description languages that allows the realization of vulnerable environments. Still, these approaches are commonly focused on the generation of the network stack and the virtual machines' deployment and do not take into account the automatic deployment and configuration of the application layer stack. Nowadays, many attacks exploit the public company website, which is commonly developed by using Content Management Systems (CMS), i.e., frameworks that allow the management and the creation of digital content for websites. According to Iqbal et al. (2020)~\cite{Iqbal2020ANES}, WordPress is the used most widely  adopted in the market (see Table~\ref{tab:marketshare}). 
\begin{table}[htb!]
\caption{Market share of CMS Platforms}
\label{tab:marketshare}
\centering
\begin{tabular}{|c|c|c|c|}
\hline
\textbf{Time} & \textbf{WordPress} & \textbf{Joomla} & \textbf{Drupal} \\ \hline
1 Apr 2019    & 58.8 \%             & 7.0\%           & 4.7\%           \\ \hline
1 Aug 2019    & 59.3\%             & 6.8\%           & 4.7\%           \\ \hline
1 Dec 2019    & 59.9\%             & 6.6\%           & 4.7\%           \\ \hline
10 Mar 2020   & 60.1\%             & 6.2\%           & 4.2\%           \\ \hline
\end{tabular}
\end{table}

In this work, we propose an approach to automate the creation of vulnerable WordPress environments through lightweight containers. Containers are environments for executing processes with configurable isolation and resource limitations~\cite{Opencontainers}. The container-based virtualization leverages several Linux kernel features, such as namespaces, resource limits, and mounts, to increase performance and optimize overall resource utilization. Such features prove very useful when developing vulnerable assets that can be deployed in cyber ranges.
We show how it is possible to automatically generate vulnerable WordPress environments by thoroughly analyzing well-documented public exploits collected into exploit databases. Exploit databases, such as Exploit-DB (EDB)~\cite{offensiveSecurity}, CxSecurity~\cite{cxSecurity} and Rapid7~\cite{rapid7}, collect the so-called \emph{Proof of Concepts} (PoC) for exploiting a vulnerability, as well as several helpful information to let users understand the required preconditions that make the system vulnerable. This exploit information can be used to create a vulnerable application without any user interaction. 
In this work, we use EDB, as it contains information relevant to reaching our goal. Indeed, exploits in EDB provide useful data, such as the vulnerability description, the exploit code (or the sequence of steps needed to reproduce the exploit), and sometimes even the vulnerable component itself. Our approach is focused on web exploits regarding one of the most widely deployed CMS (Content Management System), namely WordPress~\cite{automattic}. 

This assumption is not very limiting since the heterogeneity of web exploits, such as SQL Injection, Cross-site Scripting, and Cross-site Request Forgery allows for generating a wide number of vulnerable environments. Of course, the proposed approach can be extended to generate vulnerable scenarios from any type of exploit.

The vulnerable container generation starts by selecting a public WordPress exploit present on ExploitDB. The exploit title and the metadata are analyzed and converted into helpful information to find a Docker image that satisfies the exploit preconditions. The search is performed on Docker Hub, the world's largest repository of Docker images. When an image is found, it is used as the base for the generation of a Dockerfile containing all the instructions needed in order to setup the vulnerable environment. The process is completed by installing all the components required to reproduce the vulnerability, as well as the required CMS plugins and themes.

To build these environments, we decided to use Docker~\cite{docker} due to the isolation, reproduction, and performance benefits~\cite{nakata2021cyexec} that it naturally brings. The implementation of the proposed solution is publicly accessible on GitHub~\cite{exploitDB2docker}. 

The paper is divided into six sections. Section~\ref{relWork} describes the current approaches for building cyber range environments. In Section~\ref{Design} and Section~\ref{Implementation} we discuss, respectively, the design and the implementation of our automated solution for the generation of vulnerable WordPress scenarios. 
In Section~\ref{Evaluation} we analyze how many vulnerable configurations it is possible to generate with the proposed approach automatically. Finally, in Section~\ref{sec:Conclusions} we critically discuss the obtained results, identify ways for improving the performance of our platform in terms of exploits coverage level and highlight the main directions of our future work in the field.

\section{Related Work}
\label{relWork}
Cyber ranges are widely used for many purposes, and each solution uses a different way to deploy and generate vulnerable scenarios.  
One of the most famous commercial cyber ranges is Hack The Box (HTB)~\cite{hackthebox}. HTB offers several hand-made vulnerable machines deployed remotely. Currently, HTB does not implement any solution to automate the generation of vulnerable scenarios. In fact, the owners of the platform offer a reward to third parties who act as creators of vulnerable machines. 

In contrast to HTB, we proposed the Docker Security Playground (DSP)\cite{8169747}, a self-hosted cyber range. The local installation is possible because DSP reduces the hardware resource utilization by deploying vulnerable scenarios in lightweight containers. DSP is open-source and has considerable community support. People can design their vulnerable scenarios and share them with the community through a pull request on the GitHub repository project\footnote{DSP Repository: \url{https://github.com/DockerSecurityPlayground/DSP}}.

Cliffe et al. (2017).~\cite{schreuders2017security} address another relevant issue regarding cyber ranges, that is, the automatic generation of vulnerable scenarios. In particular, they propose \textit{Security Scenario Generator} (SecGen), a robust framework that can build complex VMs based on randomized scenarios. These vulnerable environments are generated by using pre-built modules that can be combined together and automatically deployed on a VM using Vagrant\footnote{\url{https://www.vagrantup.com/}} and Puppet\footnote{\url{https://puppet.com/}}. 

Nakata et al. (2021)~\cite{nakata2021cyexec} use the same concept to propose CyExec, a Docker-based cyber range platform with a feature that can automatically generate vulnerable scenarios in a randomized fashion. The authors compare their approach with SecGen and show that container-based virtualization, as also demonstrated in~\cite{karagiannis2020sandboxing}, can give better performance and cyber range benefits in terms of isolation. The mentioned automatic generation approaches are different from ours because the vulnerable scenarios are created by combining pre-built modules, while we build the vulnerable application by customizing a Docker image that is looked for in the Docker Hub repository. Thus, CyExec and SecGen create more complicated vulnerable environments, but our strategy is more general, and automation is, in our case, achieved with minimal implementation effort.

Costa et al. (2020)~\cite{costa2020automating} propose a framework for automating the definition and deployment of arbitrarily complex cyber range scenarios. This work relies on the Virtual Scenario Description Language (VDSL), which provides a high-level specification of the scenario properties in terms of involved networks, hardware, software, CVEs (Common Vulnerability Exposure), and other related information. The VSDL model contains information to build the vulnerable machine on top of Openstack using tools like Terraform and Packer. 

A similar automated approach is proposed by Russo et al.~\cite{russo2020building}. This work presents CRACK, a framework to automatically deploy all the components of a cyber range described using a declarative language. In particular, CRACK SDL is a language based on TOSCA that allows the specification and the inter-operation of all scenario elements.

Unfortunately, both in~\cite{costa2020automating} and in~\cite{russo2020building}, it is difficult to share one or more key components of a specific training scenario with the community. Russo et al.~\cite{bernardinetti2021nautilus} have addressed this problem and proposed a solution with Nautilus. This cyber range provides a training environment and a marketplace for sharing vulnerable scenarios. Through a specific scenario specification language, Nautilus allows to semi-automate the configuration and deployment of virtualized vulnerable networks and systems. Our approach differs from the ones mentioned above since it does not require any specification document by the user. Indeed, in our case, the vulnerable environment is generated by automatically extracting information from public exploits.

Gustafsson et al.~\cite{gustafsson2020cyber} analyze the automation features in CRATE, a cyber range used by the Swedish Defense Research Agency (FOI). One of these features deals with the automatic deployment of virtual machines and their services through a JSON file containing configuration parameters and required commands to inject the vulnerability. In~\cite{kahlstrom2021automating}, Gustafsson et al. also investigate the generation of a vulnerable machine using Common Platform Enumeration (CPE) and packet managers. Indeed, CPE makes it possible to identify the versions of vulnerable software affected by a particular CVE. At the same time, packet managers streamline the installation of many applications without any user interaction. Unfortunately, CPE only takes into account the vulnerable application without considering the various dependencies that may make that specific software version vulnerable.
For this reason, to generate the desired vulnerable environment, it is necessary to combine information from both  CVE and CPE. However, automating this strategy is complex, as CVE descriptions do not have a standard structure. Instead, our approach takes advantage of the structured information contained in the exploits, defined and verified by the exploit database maintainers.

Content Management Systems (CMS) play an important role in realizing web applications. Martinez-Caro et al.~\cite{info9020027} explore the advantages and drawbacks of these solutions and show how relevant it is the security impact of using them without proper controls. 
Even if our focus is the generation of vulnerable applications, we implemented a Docker Hub API that is able to find CMS Docker images and a CMS API that allows to setup CMS applications with vulnerable plugins and themes. These APIs can be reused to generate custom CMS configurations for any development testing purpose.
The current interest in CMS security is confirmed by Seelen et al.~\cite{jagamogan2021review}. The authors propose a review of penetration testing approaches and tools used against CMS platforms. Even if the automatic generation of vulnerable environments is interesting for cyber range developers, our approach might also be used to build an evaluation platform, and we hope that this work will help security researchers verify the suitability of their own assessment tools.

\section{Design}
\label{Design}
In this section, we describe the design of ExploitWP2Docker, a platform that ``translates'' exploit descriptions into vulnerable Wordpress containers.
Automating the translation involves two steps:
\begin{itemize}
    \item[-] analyze the \textit{Proof of Concept} of the exploit to gather information about the components required to create an environment that enables the vulnerability, such as software versions, add-ons, and configurations;
    \item[-] gather all the required components to generate a vulnerable container-based stack.
\end{itemize}

The Offensive Security Exploit Database (ExploitDB) is one of the primary sources for retrieving information about public exploits. During the development of our solution, we analyzed the structure of the exploits to understand if it was possible to extract data useful for the automatic generation of the vulnerable stack.

Public exploits, in many cases, contain information that allows reproducing the vulnerability.

\subsection{ExploitDB}
\label{ExploitDB}

Each exploit in ExploitDB has a unique identifier (EDB-ID), which is also part of the URL that references the exploit web page. Exploit web pages can be divided into three sections: 

\begin{itemize}
    \item the first section contains a short title that summarizes the product, the vulnerable version, and the attack type;
    \item the second section contains metadata of the exploit, such as the author, the vulnerability type, and the publication date;
    \item the third section presents the demonstration of the exploit, also known as PoC (Proof of Concept), and a header containing valuable information written in a standard form, according to the guidelines provided by ExploitDB mantainers\footnote{Submit Report ExploitDB page: \url{https://www.exploit-db.com/submit}}.
\end{itemize}

As it is possible to observe, several relevant bits of information can be used to automate the generation of vulnerable stacks. 

Even if this approach seems to be relatively promising, we will observe that there are many limitations in terms of vulnerability generation, even when only focusing on CMS platforms.

\subsection{WordPress exploits}
\label{WordPressExploit}

WordPress is a Content Management System (CMS) designed to be flexible. It is used to create web applications, and it is highly customizable by installing add-ons such as \textit{themes} and \textit{plugins}. Extensions can be installed through a graphical interface or via a command-line. 
By analyzing the first section of WordPress exploits in ExploitDB, we observed that a significant number of them uses a precise title pattern:
\begin{center}
   {\small WordPress $\langle$ \textit{Core/Plugin/Theme} $\rangle$ [\textit{Product}] [\textit{Version}] - [\textit{Attack Type}]}
\end{center}

The \texttt{$\langle$ Core/Plugin/Theme $\rangle$} field identifies the ex\-ploit category, i.e., details whether the vulnerability af\-fects the WordPress Core or its extensions. \texttt{[Product]} and \texttt{[Version]} fields are optional and describe the product na\-me and the CMS version.

To evaluate the effectiveness of our analysis, Table~\ref{tab:stats-wordpress} reports how many exploits follow the title pattern. In particular, the last row shows that only $1.33$\% of the exploit titles do not follow the identified pattern. For these exploits, the current  ExploitWP2Docker implementation is not able to generate a vulnerable environment.

\begin{table}[!t]
\renewcommand{\arraystretch}{1.3}
\caption{WordPress exploits statistics}
\label{tab:stats-wordpress}
\centering
\begin{tabular}{|c||c|}
\hline
WordPress       & N. of  exploit \\ \hline
Core            & 90             \\ \hline
Theme           & 1167           \\ \hline
Plugin          & 79             \\ \hline
Not Categorized & 18             \\ \hline
\end{tabular}
\end{table}
It is possible to parse the title from each WordPress exploit to obtain information regarding the vulnerable CMS version and build a vulnerable WordPress core stack.

To generate a vulnerable environment, ExploitWP2Docker searches on Docker Hub a WordPress Docker image containing a valid vulnerable tag version. 
As the title present in ExploitDB does not explicitly mention the specific WordPress version, we implement an API that queries the CPE Dictionary~\cite{nvd} provided by NIST to convert CVE information into CPE strings, in the same way as the one proposed by Gustafsson et al.~\cite{kahlstrom2021automating}. CPE strings are then parsed to discover the right vulnerable version. 

\section{Implementation}
\label{Implementation}

This section describes the implementation of the container generation process. 
In particular, in the following, we explain the environment generation procedure to build vulnerable WordPress environments\footnote{It is possible to find a sequence diagram for the WordPress vulnerable container generation process at the following URL: \url{https://github.com/NS-unina/cve2docker/blob/main/docs/gen-config-wp-flowchart.png}}.

We also analyze when the proposed approach is not able to generate a vulnerable working configuration. 
\begin{figure}[t!]
\begin{algorithmic}
\Procedure{WordPressExploit}{EID}
\State Type = ExtractTypeFromExploitTitle(EID);
\State Target = ExtractTargetFromExploitTitle(EID);
\If{Target.Version != NULL}
    \State Version = ExtractVersionFromTarget(EID);
\ElsIf{ExtractVersionFromPoC(EID)}
    \State Version = ExtractVersionFromPoC(EID);
\ElsIf{!CheckVulnAppFromExploit(EID)}
    \State raise Exception("No Vulnerable Application")    
\EndIf
\If{Type == Core}
    \State SearchWordPressDockerImage(Version); 
    \If{!SearchWordPressDockerImage(Version)}
        \State raise Exception("No Image")    
    \EndIf
    \ElsIf{CheckSoftwareLink(EID)}
        \State Product = ExtractProductFromSoftwareLink(EID);
    \Else
        \State Product = ExtractProductFromTarget(EID);
    \EndIf
    \If{Type == Plugin}
        \State CheckoutPluginFromSVNRepository(Product);
    \Else
        \State CheckoutThemeFromSVNRepository(Product)
    \EndIf
    \If{!Checkout()}
        \If{CheckSoftwareLink(EID)} 
            \State DownloadFromSoftwareLink(EID);
        \ElsIf{CheckVulnApp(EID)}
            \State DownloadFromExploitDB(EID);
        \Else  
            \State raise Exception("No Vulnerable Application")    
        \EndIf
    \EndIf
\If{!SetupConfiguration()}
    \State raise Exception("Error During Setup")      
\EndIf
\State \textbf{return} 0
\EndProcedure
\end{algorithmic}
\caption{WordPress Exploit}\label{fig:wordpress-exploit}
\end{figure}

When a user wants to generate a vulnerable environment from a WordPress exploit, ExploitWP2Docker tries to extract the vulnerable software version either from the title of the exploit or by parsing the Proof of Concept text. If it is not able to find the version, ExploitWP2Docker verifies the presence of a vulnerable application related to the exploit in ExploitDB.
Next, the procedure verifies if the vulnerability affects the WordPress Core. In this case, ExploitWP2Docker searches for a vulnerable WordPress Docker Image on Docker Hub. 
If the image is not present, the software raises an exception. Otherwise, the program is ready to generate the vulnerable environment and run the setup phase explained further below.

If the vulnerability is related to WordPress extensions, the process is different since the vulnerable Plugin or Theme source code is required to generate the vulnerable environment. WordPress extensions can be found in several ways: 

\begin{itemize}
    \item by exploring the WordPress SVN, a public collection of WordPress Themes\footnote{SVN Theme:\url{https://themes.svn.wordpress.org/}} and Plugins\footnote{SVN  Plugin:\url{https://plugins.svn.wordpress.org/}} managed via \emph{Subversion}, a well-known version control system.
    \item by parsing software links in the Proof Of Concept header.
    \item by leveraging the vulnerable applications provided by ExploitDB.
\end{itemize}

ExploitWP2Docker throws an exception if it is not able to retrieve the plugin or theme automatically.
Once all the components are obtained, the environment is configured, and, as soon the container terminates the bootstrap operation, ExploitWP2Docker executes a setup script that completes the configuration. 
Eventually, the generated stack starts and is initialized through a bootstrap phase. The bootstrap operation is completed when the WordPress setup page (\url{http://localhost/wp-admin/index.php}) is available. ExploitWP2Docker makes HTTP requests every $10$ seconds to such a URL until it receives a $200$ HTTP response code. A timeout is set to check if the services do not bootstrap correctly. 

When the bootstrap phase is completed, the setup script is executed, and, if necessary, the vulnerable plugins and themes are enabled.

The algorithm in Figure~\ref{fig:wordpress-exploit} shows the execution flow. The project has been implemented in Java by using the Spring framework. At the time of writing (i.e., software version 0.0.3), it is composed of six Java packages, 37 Java classes, and 2587 lines of code. The project has been designed to be highly modular, extendible, and configurable. In fact, several service classes have already been designed, such as \texttt{JoomlaService} and \texttt{PhpWebAppService}, that can be used to generate vulnerable environments for Joomla and PHP stacks. Anyway, such classes are in the development stage, and further works should be performed to increase the vulnerability coverage for these stacks.   

\section{Evaluation}
\label{Evaluation}
\begin{figure}[t!]
    \centering
    \centerline{\includegraphics[width=\columnwidth]{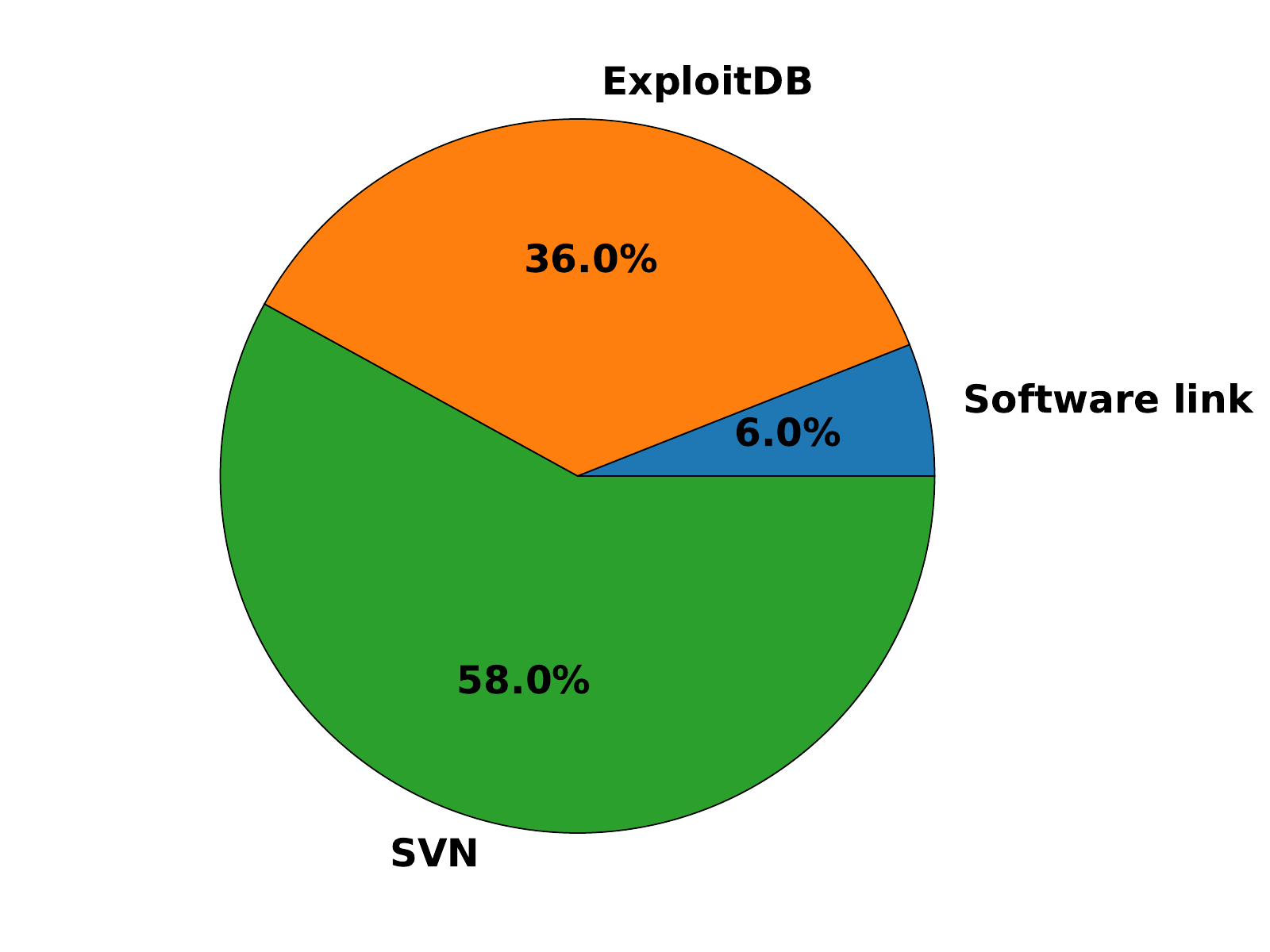}}
    \caption{Number of vulnerable downloaded applications for site - WordPress}
    \label{fig:wordpress-pie-chart-download}
\end{figure}

\begin{table*}[htb!]
\caption{Comparison with other platforms}
\label{tab:comparison}
\def\arraystretch{1.5}
\begin{tabularx}{\textwidth}{|Y|Y|Y|Y|Y|Y|Y|Y|Y|Y|}
\hline
\textbf{Work}    & \textbf{Creation of complex scenarios} & \textbf{IaC automatically generated} & \textbf{Web Server Installation} & \textbf{Web Server Configuration} & \textbf{MySQL Installation} & \textbf{MySQL Configuration} & \textbf{WP Installation} & \textbf{WP Version Configuration} & \textbf{WP Plugin Installation} \\  \hline
VSDL \cite{costa2020automating}          & \cmark  & \xmark                                                         & \cmark                                &\cmark                                  & \cmark                            & \xmark                            & \xmark                        & \xmark                                 & \xmark                               \\ \hline

\hline
CRACK \cite{russo2020building}         & \cmark  & \xmark & \cmark  & \cmark  & \cmark  & \cmark  & \cmark  & \xmark & \xmark                               \\ \hline
Nautilus  \cite{bernardinetti2021nautilus}       & \cmark  & \xmark                                                         & \cmark                                & \cmark                                 & \cmark                           & \cmark                            & \xmark                        & \xmark                                 & \xmark                               \\ \hline
ExploitWP2 Docker & \xmark  & \cmark                                                         & \cmark                                & \cmark                                 & \cmark                           & \cmark                            & \cmark                        & \cmark  & \cmark \\ \hline
\end{tabularx}
\end{table*}
To the best of our knowledge, no works address the automation of generating vulnerable web applications for cyber ranges. For this reason, it is not possible to compare the effectiveness of the proposed approach with other ones. Table~\ref{tab:comparison} shows a feature comparison between our work and the current state-of-the-art platforms that automate the realization of cyber-range scenarios. As it is possible to observe, all the related works entirely cover the deployment and configuration of complex scenarios, from the host configuration to the webserver. However, such solutions do not automate the generation of the source code used to build the infrastructure, i.e., Infrastructure as Code (IaC) is not automatically generated. Furthermore, they do not allow the configuration of the web application stack, i.e., it is not possible to setup, install and configure a WordPress instance. 
As already stated, this is related to the fact that related works are focused on defining semantic languages to simplify the realization of complex vulnerable environments, while our solution addresses the specific problem of realizing vulnerable applications. On the same time, our solution is not able to generate complex infrastructures, but it would be possible to integrate ExploitWP2Docker into such solutions to bring the benefits of both approaches.
The remainder of this section evaluates the number of generated configurations obtained by executing ExploitWP2Docker on all the WordPress exploits present on ExploitDB. 
The presented analysis was carried out in May $1^{st}$ $2021$.
We were able to successfully generate\textbf{ 484 scenarios}, i.e.,  39\% of the total number of WordPress exploits.Fig.~\ref{fig:wordpress-pie-chart-download} illustrates the main source archives leveraged by ExploitWP2Docker to download the extensions required for building the vulnerable environment. Nearly two-thirds of the extensions are obtained from the SVN repository, while $36$\% are downloaded from ExploitDB. The remaining part comes from links inside the PoC header. 
These results highlight the importance of implementing alternative approaches to download the components that build the stack of a vulnerable environment. As WordPress offers several sources, it is possible to increase automation effectiveness. However, the approach should be extended to other CMS systems, such as Joomla, that do not have available add-ons repositories. 
Nevertheless, the results obtained by the open-source scanners, namely, Arachni and ZAP, are very satisfactory and show that open-source solution performance can be compared to commercial scanners. 
The result of the work demonstrates that WAVSEP is an excellent platform for benchmarking web vulnerability detection scanners. The methodology used allowed the goals to be met by enabling a current benchmark platform and can, therefore, be used to keep the platform up-to-date and as a basis for future developments.
Moreover, the current state of the scanners is still far from the activity performed by a security expert, such as a Penetration Tester, since all test cases are vulnerable. It should be noted that although the results obtained are satisfactory, the failure to detect even a single test case can mean the presence of vulnerabilities within a web application. Scanners, both open-source and commercial, are excellent support tools. Better results could be achieved, for example, by integrating artificial intelligence techniques. 

\begin{figure}[t!]
    \centering
    \includegraphics[width=\linewidth]{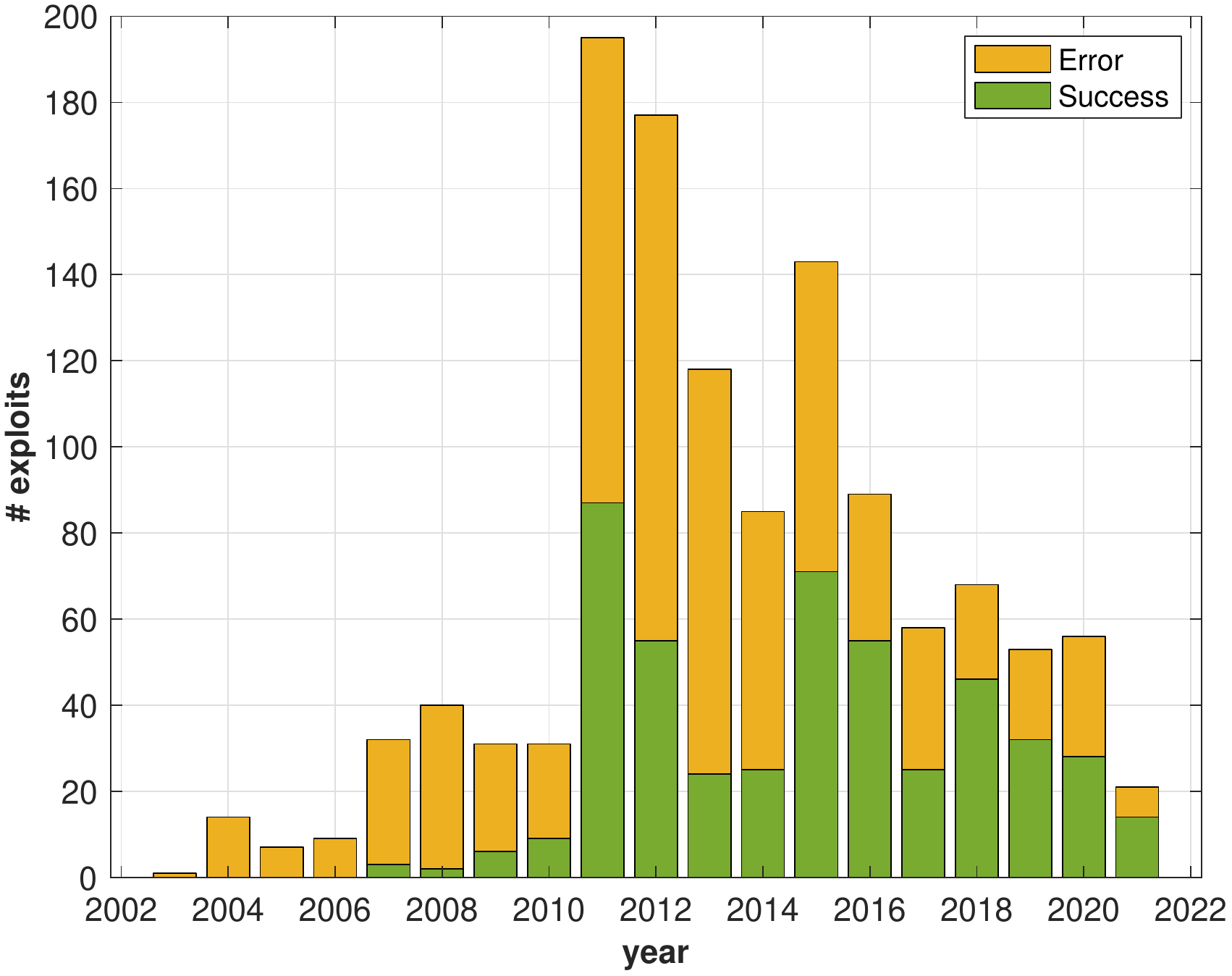}
    \caption{Number of generation divided by year - WordPress}
    \label{fig:wordpress-bar-chart-total}
\end{figure}

 Fig. \ref{fig:wordpress-bar-chart-total} illustrates the number of submitted exploits and environments correctly generated, divided by year.

\begin{figure}[t!]
    \centering
    \centerline{\includegraphics[width=\columnwidth]{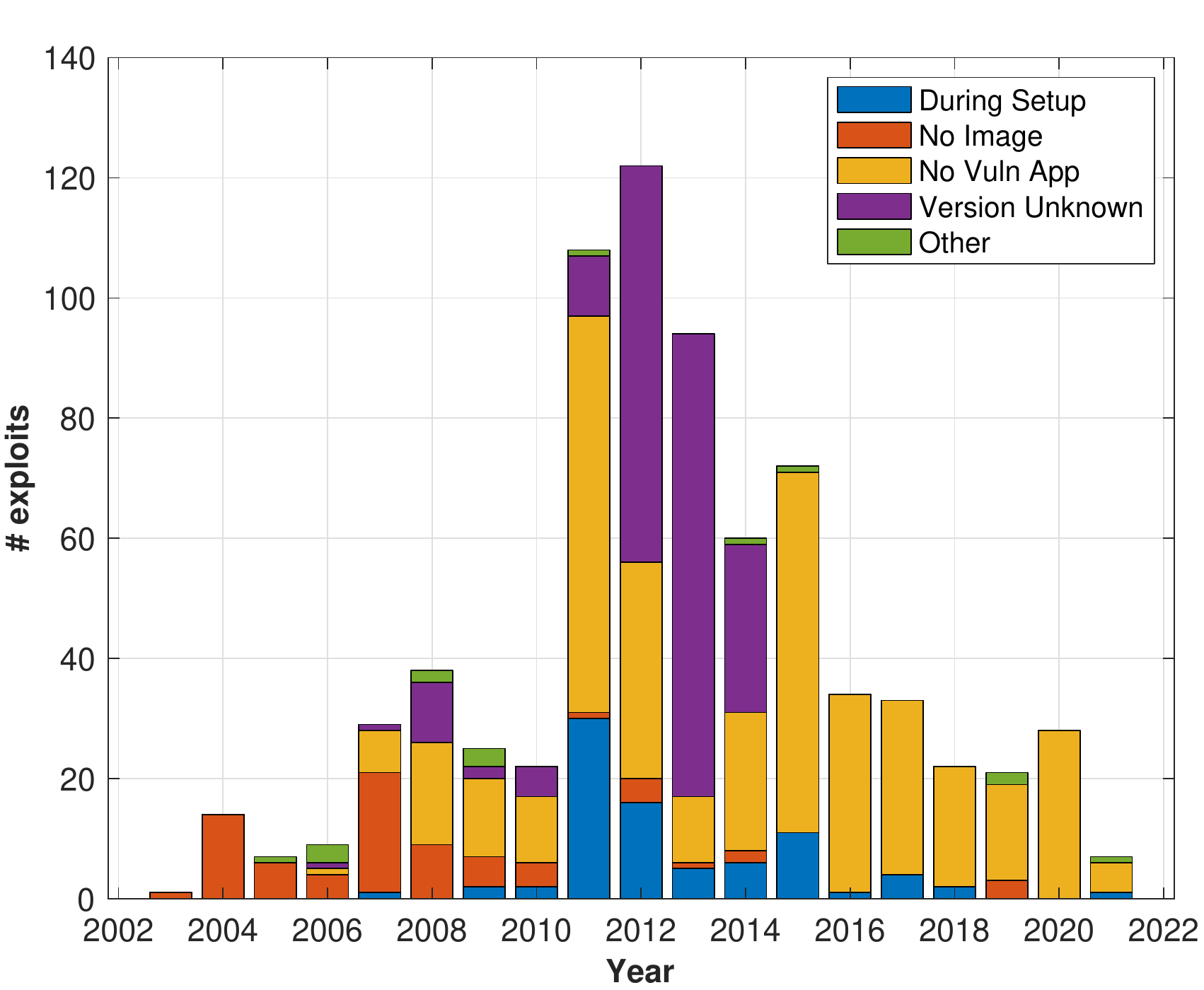}}
    \caption{Error of generations divided by year - WordPress}
    \label{fig:wordpress-bar-chart-error}
\end{figure}

In fact, as it is possible to observe in Fig.~\ref{fig:wordpress-bar-chart-error}, the vast majority of the exploits submitted between $2003$ and $2007$ are related to core WordPress vulnerabilities. In these cases, ExploitWP2Docker would need to download a vulnerable Docker image, which is unfortunately not feasible since the first WordPress version available on Docker Hub (version $3.1.0$) was released in $2011$.

Also, the approach fails when the vulnerable application is not available or the WordPress version is unknown. One other failure cause can be the incompatibility between the vulnerable plugin and the adopted WordPress version. The pattern title was adopted by ExploitDB only in recent years, so older exploits do not report any information about the vulnerable core or plugin version.
In these cases, it is not possible to use our approach.

\section{Conclusion}
\label{sec:Conclusions}

In this work, we have presented a container-based platform that automates the generation of vulnerable WordPress environments.
With our approach, we were able to build more than four hundred vulnerable WordPress environments. The number of such vulnerable environments will certainly increase over time with the release of new exploits.
Our future work will expand the focus towards other systems beyond WordPress, starting from other CMS, such as Joomla and PHP-based web applications. As a matter of fact, if we extend this approach to all possible PHP applications, we will be able to reproduce in an automated fashion about $50$\%\footnote{ExploitDB stat: \url{https://www.exploit-db.com/exploit-database-statistics}} of all the exploits contained in ExploitDB. This goal can be achieved with minimal effort, thanks to the modular setting of the system we have devised.
Relevant improvements could be achieved by realizing Docker images for CMS versions not currently available on Docker Hub. Another interesting evolution of our work deals with exploring alternative approaches to increase the coverage of extensions. Concerning this point, we do believe that the use of a headless browser will allow us to effectively reproduce the typical human interaction patterns, hence automating the download of both themes and components.
A further improvement can certainly derive from extending the proposed approach to other exploit databases.
As a final remark, we acknowledge that containers do limit the reproduction of vulnerable environments, as we depicted in~\cite{Caturano2020}.
Some authors are experimenting alternative virtualization approaches, such as using \textit{MicroVM}, which has the same benefits as container-based virtualization in terms of resource overhead but offers a better grade of isolation \cite{karagiannis2020sandboxing}. 
As part of our future work, we will explore the application of alternative virtualization techniques, such as \textit{MicroVM}, for the generation of vulnerable environments not reproducible through Docker.

\bibliographystyle{IEEEtran}
\bibliography{IEEEabrv,references}
\end{document}